\documentstyle[12pt,bezier]{article}
\newfont{\mathea}{msam10 scaled\magstep0}
\newfont{\matheb}{msbm10 scaled 1095}
\newfont{\tmpEins}{cmsy10 scaled 2074}
\newfont{\tmpZwei}{cmsy10 scaled 1095}
\newfont{\tmpDrei}{cmsy10 scaled 1000}
\newfont{\tmpVier}{cmsy5 scaled 1000}
\newfont{\tmpFuenf}{msbm7 scaled\magstep0}

\def\dach#1#2{\mbox{$\mathop{\vbox{\ialign{%
  $##\crcr\hfil #1 \hfil$\crcr}}}\limits^{\scriptscriptstyle #2}$}}

\def\rnzs{\dach{\rho_2}{\mbox{$\scriptscriptstyle\kern-.7mm0$}}\kern-1.2mm'}

\def\Subset{\mbox{$\subset\kern-.5mm\subset$}}
\newcommand{\LI}{\mbox{{\rm L$^{\kern-.15em\raise.2ex\hbox{\scriptsize 1}}$}}}

\def\Ldummy{\left.\bgroup}
\def\Rdummy{\egroup^{\rule{0mm}{1.4mm}}\right.}
\def\LA{\left\langle\bgroup}
\def\RA{\egroup^{\rule{0mm}{1.4mm}}\right\rangle_{\cal A}^{}}
\def\LR{\left(\bgroup}
\def\RR{\egroup^{\rule{0mm}{1.4mm}}\right)}
\def\LG{\left\{\bgroup}
\def\RG{\egroup^{\rule{0mm}{1.4mm}}\right\}}

\def\Wort#1{\mbox{{\rm #1\kern.1em}}}

\def\lfac#1#2{\vcenter{\hbox{$#1\kern-.2em\raise-.6ex\hbox{\Large{/}}%
 \kern-.2em\raise-1.2ex\hbox{$#2$}$}}}

\def\gin{\mbox{\tmpZwei\symbol{91}\kern-1.4mm\rule{.2mm}{1.85mm}\kern1.4mm}}
\def\gni{\mbox{\tmpZwei\symbol{92}\kern-1.4mm\rule[.15mm]{.2mm}{1.85mm}%
  \kern1.4mm}}

\def\EINS{{\mathop{1\kern-.25em\mbox{{\rm{\small l}}}}}}

\begin{document}
\large Friedmann equation and SCP-results 

\vspace{5mm}

\normalsize Hellmut Baumg\"artel

Mathematical Institute

University of Potsdam

Germany

e-mail: baumg@uni-potsdam.de

\begin{abstract}
This note emphasizes that only special solutions of the Friedmann equation 
are compatible with the results of the "Supernovae Cosmology Project" (SCP).
The curvature parameter of these solutions equals +1 and
there is a simple relation between the minimal Hubble
parameter, the cosmological constant and the total mass.
\end{abstract}

\section{Introduction}

Obviously the results of the SCP support the conclusion that in the course of the time
evolution of the universe the Hubble parameter must have passed in the past
through a minimum, because
in the deceleration epoch it was decreasing whereas the results of SCP show that there
is a time where it is increasing (see e.g. [1,2,3]).

\vspace{3mm}

Taking the Friedmann equation as a basis for the time evolution (neglecting first radiation's
contribution) it is known that there is only one type of solutions of this equation where
the Hubble parameter assumes a minimum value (see e.g. [4, p.417]). 
For this type the curvature parameter equals +1
and $D>0$ where D is the discriminant of the equation of third degree obtained by 
multiplication
of the right hand side of the Friedmann equation by R. This means that only the spherical
ansatz for the basic Robertson-Walker metric is compatible with the observations of SCP. 
In this case the density of matter is $M/2\pi^{2}R^{3}$ where $M$ is the total mass. 
The R-value for the mentioned minimum can be calculated explicitly. The condition $D>0$ 
yields a lower estimate for the cosmological constant and there is a simple relation
between the minimal Hubble parameter, the cosmological constant and the total mass.

Taking into account radiation's contribution these results remain true, the R-value
for the minimum of the Hubble parameter undergoes a small correction but it can be
calculated explicitly, too.

\section{The Friedmann equation}

Without taking into account radiation's contribution the Friedmann equation reads
\begin{equation}
\left(\frac{dR}{dt}\right)^{2}=\frac{\alpha}{R}+\beta R^{2}-\gamma,\quad R>0,
\end{equation}
with
\begin{equation}
\alpha:=\frac{8\pi G}{3}A,\quad \beta:=\frac{1}{3}\Lambda c^{2},\quad \gamma:=\epsilon c^{2},
\end{equation}
where 
$A:=\rho(t)R(t)^{3}$ is a constant ($\rho$ is the density),
$\Lambda$ denotes the cosmological constant, $c$ the velocity of light, $G$ the
gravitational constant and $\epsilon:=0,\pm 1$ the curvature parameter. The Hubble
parameter is defined by
\[
H:=\frac{R'}{R}.
\]
In the following it is assumed that $\Lambda>0$.

\vspace{3mm}

The structure of the solutions $R(\cdot)$ of (1) is determined by the properties of the
roots of the equation of the third degree
\begin{equation}
R^{3}+pR+q=0,\quad p:=-\frac{\gamma}{\beta},\quad q:=\frac{\alpha}{\beta}
\end{equation}
with discriminant
\[
D:=\left(\frac{\alpha}{2\beta}\right)^{2}-\left(\frac{\gamma}{3\beta}\right)^{3}.
\]
To obtain an overview on the different types of solutions of (1) one distinguishes
several cases (see e.g. [4, p. 417 ff.]).

\vspace{3mm}

Case I. $\epsilon=0$ or $\epsilon=-1$.

Then necessarily
\[
D>0
\]
follows, i.e. in this case the equation (3) has one negative real root and two complex-
conjugated roots. Then
\[
R^{3}+pR+q>0,\quad R>0,
\]
there is no restriction for $R$. Furthermore
\[
H^{2}=\beta+\frac{\alpha}{R^{3}}-\frac{\gamma}{R^{2}}
\]
and this implies that $H(t)$ is strongly monotonically decreasing, for $R'(t)>0$
as well as for $R'(t)<0$.

\vspace{3mm}

Case II. $\epsilon=+1$.

In this case one has to distinguish between the subcases $D>0,\;D<0$ and $D=0$. The latter
case is omitted (it is the branching case for solutions of (1)).

Case II(i). $D<0$.

This means $27\alpha^{2}\beta<4\gamma^{3}$. In this case the equation (3) has three
real roots, $r_{0}<0,\;0<r_{1}<r_{2}$, where
\[
r_{2}:=2\left(\frac{\gamma}{3\beta}\right)^{1/2}\cos\frac{\phi}{3},\quad
r_{1}:=\left(\frac{\gamma}{3\beta}\right)^{1/2}\left(\sqrt{3}\sin\frac{\phi}{3}-
\cos\frac{\phi}{3}\right)
\]
and
\[
cos\phi=-\frac{\alpha}{2\beta}\left(\frac{\gamma}{3\beta}\right)^{-3/2},\quad
\frac{\pi}{2}<\phi<\pi,\quad\mbox{i.e.}\quad \frac{1}{2}<\cos\frac{\phi}{3}<\frac{1}{2}
\sqrt{3}.
\]
This means the interval $r_{1}<R<r_{2}$ is not allowed for solutions of (1), there are
two admissible regions, $R>r_{2}$ and $0<R<r_{1}$. Note that in this case the numbers
$\frac{3\alpha}{2\gamma}$ and $\left(\frac{\alpha}{2\beta}\right)^{1/3}$ lie in the
forbidden interval,
\[
r_{1}<\frac{3\alpha}{2\gamma}<\left(\frac{\alpha}{2\beta}\right)^{1/3}<r_{2}.
\]
This implies that in the case $0<R(t)<r_{1}$ the function $H(\cdot)$ is strongly
monotonically decreasing and in the other case $R(t)>r_{2}$ it is strongly monotonically
increasing.

\vspace{3mm}

Case II(ii). $D>0$

This means 
\begin{equation}
27\alpha^{2}\beta>4\gamma^{3}.
\end{equation} 
In this case the equation (3) has one negative
real root $-r_{0},r_{0}>0,$ and two complex-conjugated roots $x_{0}\pm iy_{0},y_{0}>0$,
hence
\[
R'(t)^{2}=\frac{\beta}{R}(R+r_{0})((R-x_{0})^{2}+y_{0}^{2})\geq\beta y_{0}^{2},
\quad R>0,
\]
i.e. either 
\begin{equation}
R'(t)\geq\sqrt{\beta}\,y_{0},
\end{equation}
hence $R(\cdot)$ is strongly monotonically increasing
or $R'(t)\leq -\sqrt{\beta}\,y_{0}$ and this means that $R(\cdot)$ is strongly monotonically
decreasing. Only the first case is considered. In this case one has
\begin{equation}
\lim_{t\rightarrow\infty}R(t)=\infty.
\end{equation}
Furthermore,
\[
\left(\frac{\alpha}{2\beta}\right)^{1/3}<\frac{3\alpha}{2\gamma}
\]
and a simple calculation yields
\[
R_{w}:=\left(\frac{\alpha}{2\beta}\right)^{1/3}\quad\mbox{is a turning point for}\quad R(\cdot).
\]
From
\begin{equation}
H(t)^{2}=\beta+\frac{\alpha}{R^{3}}-\frac{\gamma}{R^{2}}
\end{equation}
one obtains
\[
H'(t)=-\frac{3\alpha}{2}\frac{1}{R^{3}}+\gamma\frac{1}{R^{2}},
\]
i.e. $R_{0}:=\frac{3\alpha}{2\gamma}$ is a zero of $H'(\cdot)$. 
The corresponding time is denoted by $t_{0}$. Since
\[
H''(t_{0})=\frac{3\alpha}{2}R'(t_{0})\frac{1}{R_{0}^{4}}>0,
\]
this means that 
\[
R_{min}:=\frac{3\alpha}{2\gamma}\quad \mbox{is a minimum for} \quad
H(\cdot)
\]
and $R_{w}<R_{min}$. Moreover,
\[
R_{w,H}:=\frac{9\alpha}{4\gamma}> R_{min}\quad\mbox{is a turning point for}\quad H(\cdot)
\]
Finally, the minimal value of $H(t)^{2}$ equals
\begin{equation}
H(t_{min})^{2}=\beta-\frac{4\gamma^{3}}{27\alpha^{2}}.
\end{equation}
From (7) one obtains
\[
H(\infty)=\sqrt{\beta},
\]
this is the asymptotic value for $H(\cdot)$. The value of $H(\cdot)$ at the turning point
$R_{w}$ of $R(\cdot)$ is given by
\begin{equation}
H(t_{w})^{2}=3\beta-\gamma\left(\frac{\alpha}{2\beta}\right)^{-2/3}
\end{equation}
Note that the turning point for $R(\cdot)$ does not depend on $\gamma$. For example, if
$\alpha>2\beta^{5/2}$ then $R_{w}>\sqrt{\beta}$ follows. The position of $H(t_{w})$ w.r.t.
$\sqrt{\beta}$ is simple, too: $H(t_{w})>\sqrt{\beta}$ if $8\alpha^{2}\beta>4\gamma^{2}$ 
and $H(t_{w})<\sqrt{\beta}$ if $8\alpha^{2}\beta<4\gamma^{2}$. Note (4) in this connection.
The relation between $R(t_{w})=R_{w}$ and $H(t_{w})$ is more complicated. One has to compare
the right hand side of (9) with $R_{w}$.

\section{Conclusion}
The case II is the spherical case for the solutions of (1), the volume of the 3-sphere is
$V(t)=2\pi^{2}R(t)^{3}$ and the constant $A$ in (2) is given by $A:=M/2\pi^{2}$, where $M$
is the total mass.

Only the solutions of (1) in the case II(ii) have the property that the corresponding 
Hubble parameter has a minimum, i.e. the time before is the "decelerating epoch", the 
time after an "accelerating epoch" and the turning point $9\alpha/4\gamma$ of the Hubble 
parameter reflects the weakening of this acceleration with the final convergence point
$\sqrt{\beta}$.
 
For the special values (2) one obtains the following expressions: First $D>0$ implies
\[
\Lambda> \frac{\pi^{2}}{4}\frac{c^{4}}{(GM)^{2}}.
\]
The minimal $R$ for the Hubble parameter is given by
\[
R_{min}=\frac{2}{\pi}\frac{GM}{c^{2}}
\]
and the minimal value of the Hubble parameter by
\[
H(t_{min})^{2}=\frac{1}{3}c^{2}\left(\Lambda-\frac{\pi{2}}{4}\frac{c^{4}}{(GM)^{2}}\right),
\]
i.e. from the minimal value of the Hubble parameter and the total mass  one is able to 
calculate the cosmological constant.

\section{References}
[1] S. Perlmutter: Supernovae, Dark Energy, and the Accelerating Universe, Physics Today,
April 2003, 53-60

\noindent [2] P. Astier, J. Gay, N. Regnault et al.: Astronomy and Astrophysics 447, 31 (2006)

\noindent [3] B. Leibundgut: Creating a legacy and learning about dark energy, Astronomy
and Astrophysics 500, 615-616 (2009)

\noindent [4] R. Tolman: Relativity, Thermodynamics and Cosmology, Oxford 1934, Russian
translation Moscow 1974

\end{document}